\documentclass[aps,prd,superscriptaddress,floatfix,nofootinbib,preprintnumbers,eqsecnum,twocolumn]{revtex4-1}

\pdfoutput=1
\usepackage{amssymb,amsfonts,amsmath,graphicx}
\usepackage{xcolor}
\usepackage{orcidlink}
\usepackage[english]{babel}
\usepackage[T1]{fontenc}
\usepackage{amsmath}
\usepackage{slashed}
\usepackage{booktabs}
\usepackage{listings}
\usepackage[utf8]{inputenc}
\usepackage{hyperref}
 \hypersetup{ 
   colorlinks,
   linkcolor={blue!80!black},
   citecolor={blue!70!black},
   urlcolor={blue!70!black}
 }
 
\usepackage{placeins}

\usepackage{amssymb,amsmath,epsfig,todonotes,longtable}

\numberwithin{equation}{section}

\newcommand{\bean}{\begin{eqnarray*}}
\newcommand{\eean}{\end{eqnarray*}}

\newcommand{\fref}[1]{Figure~\ref{#1}}

\newcommand{\rk}{\mathop{{\rm rk}}}

\newcommand{\ind}{\mathop{{\rm ind}}}

\newcommand{\IP}{\mathbb{P}}

\newcommand{\cN}{{\cal N}}

\newcommand{\cA}{{\cal A}}
\newcommand{\cB}{{\cal B}}
\newcommand{\cC}{{\cal C}}

\newcommand{\cK}{{\cal K}}
\newcommand{\cV}{{\cal V}}

\def\cjn1{{\cA, \cC^*\otimes \wedge^j \cN^*}}
\def\bjn1{{\cA, \cB^*\otimes \wedge^j \cN^*}}
\def\vjn1{{\cA, \cV^*\otimes \wedge^j \cN^*}}
\def\cjn2{{\cA, \cC\otimes \wedge^j \cN^*}}
\def\bjn2{{\cA, \cB\otimes \wedge^j \cN^*}}
\def\vjn2{{\cA, \cV\otimes \wedge^j \cN^*}}

\newcommand{\cicy}[2]{\begin{matrix} #1\end{matrix}\!\left[\begin{matrix}#2 \end{matrix}\right]}

\newcommand{\be}{\begin{equation}}
\newcommand{\ee}{\end{equation}}
\newcommand*{\nnbe}{\begin{equation}}
\newcommand*{\nnee}{\end{equation}}
\newcommand{\bea}{\begin{eqnarray}}
\newcommand{\eea}{\end{eqnarray}}
\newcommand{\ba}{\begin{align}}
\newcommand{\ea}{\end{align}}
\newcommand{\bi}{\begin{itemize}}
\newcommand{\ei}{\end{itemize}}

\newsavebox{\overlongequation}

\begin{document}
\title{Decoding Nature with Nature's Tools: \\ Heterotic Line Bundle Models of Particle Physics with Genetic Algorithms and Quantum Annealing}



\author{Steve A.~Abel \orcidlink{0000-0003-1213-907X}}
\email[]{s.a.abel@durham.ac.uk}
\affiliation{IPPP, Durham University, Durham, DH1 3LE UK\\ Department of Mathematical Sciences, Durham University,
Durham DH1 3LE, UK}
\affiliation{CERN, Theoretical Physics Department, CH 1211 Geneva 23 Switzerland}

\author{Andrei Constantin \orcidlink{0000-0002-0861-5363}}
\email[]{andrei.constantin@physics.ox.ac.uk}
\affiliation{Rudolf Peierls Centre for Theoretical Physics, University of Oxford, Parks Road, Oxford OX1 3PU, UK}

\author{Thomas R.~Harvey \orcidlink{0000-0002-4990-4778}}
\email[]{thomas.harvey@physics.ox.ac.uk}
\affiliation{Rudolf Peierls Centre for Theoretical Physics, University of Oxford, Parks Road, Oxford OX1 3PU, UK}

\author{Andre Lukas \orcidlink{0000-0003-4969-0447}}
\email[]{andre.lukas@physics.ox.ac.uk}
\affiliation{Rudolf Peierls Centre for Theoretical Physics, University of Oxford, Parks Road, Oxford OX1 3PU, UK}

\author{Luca A. Nutricati \orcidlink{0000-0002-5045-5113}}
\email[]{luca.a.nutricati@durham.ac.uk}
\affiliation{IPPP, Durham University, Durham, DH1 3LE UK\\ Department of Mathematical Sciences, Durham University,
Durham DH1 3LE, UK}


\begin{abstract}\noindent
The string theory landscape may include a multitude of ultraviolet embeddings of the Standard Model, but identifying these has proven difficult due to the enormous number of available string compactifications. Genetic Algorithms (GAs) represent a powerful class of discrete optimisation techniques that can efficiently deal with the immensity of the string landscape, especially when enhanced with input from quantum annealers. In this letter we focus on geometric compactifications of the $E_8\times E_8$ heterotic string theory compactified on smooth Calabi-Yau threefolds with Abelian bundles. We make use of analytic formulae for bundle-valued cohomology to impose the entire range of spectrum requirements, something that has not been possible so far. For manifolds with a relatively low number of K\"ahler parameters we compare the GA search results with results from previous systematic scans, showing that GAs can find nearly all the viable solutions while visiting only a tiny fraction of the solution space.  Moreover, we carry out GA searches on manifolds with a larger numbers of K\"ahler parameters where systematic searches are not feasible.
\end{abstract}

\maketitle		

\section{Introduction}
To appreciate the astounding efficiency of evolution it is useful to recall the numbers involved. The human genome contains some 3 billion base pairs; with four types of bases, the total number of potential DNA combinations reaches the unfathomably large number of $4^{3,000,000,000}$. Although it is impossible to estimate how many of these would lead to biologically functional organisms, the fraction is likely to be extremely small.    

Equally prohibitive statistics plague string theory. Estimates on the size of the string landscape (the set of mathematically consistent four-dimensional solutions of string theory) include the famous $10^{500}$ type IIB flux compactifications \cite{Douglas:2003um, Ashok:2003gk}, as well as the more recent estimate  of $10^{272,000}$ F-theory flux compactifications on a single elliptically fibered four-fold~\cite{Taylor:2015xtz}. While the number of string compactifications leading to standard-like models may in itself be as large as $10^{700}$, as estimated in Ref.~\cite{Constantin:2018xkj}, this number is nevertheless minute in comparison to the size of the entire string landscape. Random sampling is guaranteed to fail at identifying  such standard-like models from string theory, as is systematic searching which given the scales involved is simply beyond any present or future computational capabilities. Instead, one needs to employ methods as powerful as nature's.

Genetic algorithms (GAs) form a class of discrete optimisation techniques that rest on the three pillars of evolutionary dynamics: selection, breeding and mutation. While a proper mathematical framework for understanding the performance of GAs is currently lacking, the empirical evidence strongly indicates that GAs are highly efficient in identifying viable solutions within very large search spaces. In the context of string model building, `good solutions' correspond to compactifications whose low-energy symmetry and particle content match those of the Standard Model and whose (stabilised) moduli explain all its free parameters. 

In this paper we focus on geometric compactifications in the $E_8\times E_8$ heterotic string theory involving abelian bundles over smooth compact Calabi-Yau threefolds. We address the first task of recovering the gauge symmetry and the particle content of the Standard Model, while the second task of accounting for the free parameters will be addressed in a future study, also relying on GAs. The use of GAs in string phenomenology is relatively new, with several studies already indicating its huge problem-solving potential  \cite{Abel:2014xta, Halverson:2019tkf,Abel:2021rrj, Abel:2021ddu, Cole:2021nnt}. 
The present work brings three novel elements to the discussion.

Firstly, the recent discovery of analytic formulae for cohomology dimensions of line bundles over Calabi-Yau threefolds~\cite{Constantin:2018hvl, Klaewer:2018sfl, Larfors:2019sie, Brodie:2020fiq} allows us to implement a GA search with the full set of spectrum constraints. In this way we bypass computations in commutative algebra whose complexity can be as large as doubly exponential in the number of variables and which cannot possibly be carried out alongside a GA. Concretely, we search for $\mathcal N=1$ supersymmetric $SU(5)$ GUTs with four additional Green-Schwarz anomalous $U(1)$s, three $(\overline{\mathbf 5},\mathbf{10})$ generations, no exotic $\overline{\mathbf{10}}$ multiplets and at least one vector-like $\mathbf 5-\overline{\mathbf 5}$ pair to account for the Higgs fields. The presence of additional $U(1)$-factors with ultra-heavy gauge bosons implies that the low-energy gauge symmetry is enhanced by several effectively global $U(1)$s, a feature that can be efficiently exploited in the construction of improved models of Particle Physics through the constraints imposed on the allowed operators in the Lagrangian~\cite{Blumenhagen:2006wj, Blumenhagen:2006ux, Anderson:2011ns, Anderson:2012yf, Anderson:2013xka, Buchbinder:2014qda, Buchbinder:2014sya}. While not directly constructing standard-like models, past experience indicates that almost every $SU(5)$ model meeting the above requirements allows for several embeddings of the (minimally supersymmetric) Standard Model.

Secondly, for Calabi-Yau manifolds with a low number of K\"ahler parameters, we are able to confront the results of the GA search with previous systematic scans~\cite{Anderson:2013xka, Constantin:2018xkj} and reinforcement learning searches~\cite{Larfors:2020ugo}, showing that GAs are indeed capable of finding nearly all available solutions while visiting only a tiny fraction of the solution space. Moreover, we are able to probe manifolds with larger numbers of K\"ahler parameters, where systematic searches are not feasible, thereby demonstrating the real power of the method.  

Finally, following Ref.~\cite{Abel:2022bln}, we consider so-called Genetic Quantum Annealing (GQA) in which the classical GA is enhanced with input from quantum annealers. This introduces a form of direct mutation into the algorithm which, as we demonstrate, promises future significant speed-up, compared to the classical GA.


\section{Heterotic Line Bundle Models}\label{sec:Background}\label{sec:heterotic}
In String Theory the dimension of space-time is not an input, but a prediction: the internal consistency of the theory at quantum level requires this dimension to be 10. The additional six dimensions escape direct observation through compacification on manifolds of sufficiently small size, however different topologies and geometries can lead to very different four-dimensional universes. 
 
Several proposals for connecting String Theory and Particle Physics have been known since the mid-80s. Although believed to be dual to each other, each of these approaches comes with its own set of technical hurdles. The earliest and arguably one of the most promising proposal is the $E_8\times E_8$ heterotic string compactified on smooth Calabi-Yau threefolds with holomorphic vector bundles. 

In this context, the geometrical data describing the additional dimensions consists of a Calabi-Yau threefold $X$ which reduces the number of large space-time dimension from 10 to 4, and a holomorphic bundle $V$ on $X$, needed to break the $E_8\times E_8$ gauge symmetry to the Standard Model gauge group or to one of its grand unification embeddings. The set of topologically distinct pairs $(X, V)$ that can serve as compactification data is virtually unbounded, however, there are strong hints that physically viable models can only be found within a finite, though extremely large, subset~\cite{Anderson:2013xka, Buchbinder:2013dna}. 

Two key questions arise at this point. (1) How can one deal with the extremely large number of possible compactifications in order to identify the most promising ones? (2) How far can the analysis of String Theory models be pushed so as to become relevant for Particle Physics? These questions receive the best answers within the class of compactifications where $V$ is a sum of line bundles. In this case, two of the major technical difficulties, checking slope-stability of the bundle and checking the low-energy spectrum, become manageable. Stability checks are relatively straightforward due to the split nature of the bundle, while computations of the spectrum are made virtually instantaneous by the aforementioned discovery of line bundle cohomology formulae~\cite{Constantin:2018hvl, Brodie:2020fiq}. As a result, deciding the physical viability of a heterotic line bundle sum model at the level of the particle spectrum (three families of quarks and leptons, the presence of a Higgs field and the absence of any exotic matter charged under the Standard Model gauge group) can be accomplished within a fraction of a second, something that has never been possible before. By comparison, traditional constructions in the literature have taken several years of laborious work to achieve a comparable level of analysis. 

Another salient feature of line bundle models is the presence of additional $U(1)$-symmetries, which can restrict the allowed operators in the Lagrangian in a way that is robust against deformations away from line bundle sums~\cite{Buchbinder:2013dna}. In this way, the $U(1)$ symmetries give rise to Froggatt-Nielsen models of fermion masses and mixings within string theory. 

More concretely, throughout this letter, $V$ will be a rank-5 line bundle sum $V=\oplus_{a=1}^5 L_a$ over a Calabi-Yau threefold $X$, so that the resulting model has $SU(5)\times S(U(1)^5)$ symmetry. 
The notation $L_a={\cal O}_X(k_a)$ indicates a line bundle with first Chern class $c_1(L_a) = k_a^iJ_i$, where $k_a^i$ are the components of the integer vectors $k_a\in\mathbb{Z}^h$ and $(J_1,\ldots ,J_h)$ is a suitably chosen basis of $H^2(X,\mathbb Z)$, with dimension $h=h^{1,1}(X)$. The five integer vectors $(k_1,\ldots ,k_5)$ uniquely specify the line bundle sum~$V$. The manifold $X$ will be assumed to admit a free action of a non-trivial discrete group $\Gamma$, such that the quotient manifold $X/\Gamma$ has a non-trivial fundamental group (in fact, isomorphic to $\Gamma$). Given such a group action, there are, in general, several ways to break $SU(5)$ to the SM group using an appropriate discrete Wilson line on $X/\Gamma$. 
Fixing  $X$, the aim will be to identify the line bundle sums $V$ that satisfy the following constraints:
\begin{description}
\item[(C1) $E_8$ embedding]  $c_1(V) =\sum\limits_{a=1}^5 k_a\stackrel{!}{=}0$~. 
\newline 
In order to guarantee that the structure group of $V$ is $S(U(1))^5$ and not smaller, no proper subsets of line bundles in $V$ are allowed to have a vanishing first Chern class. 
\item[(C2) Anomaly cancellation] 
$$c_{2,i}(V)=-\frac{1}{2}d_{ijk}\sum_{a=1}^5 k_a^jk_a^k\stackrel{!}{\leq}c_{2,i} (TX)~,
$$
\noindent $\forall i=1,\ldots ,h$, where $d_{ijk}$ denote the triple intersection numbers and $c_2(TX)$ the second Chern class of the tangent bundle of $X$, relative to the basis $(J_1,\ldots ,J_h)$.
\item[(C3) Supersymmetry/poly-stability] There exists a non-trivial common solution $t^i$ to the vanishing slopes $$\mu(L_a)=d_{ijk}k_a^it^jt^k\stackrel{!}{=}0\text{ for }a=1,\ldots ,5$$ such that $J=t^iJ_i$ is in the interior of the K\"ahler cone, which in our examples corresponds to $t^i>0$. Solving the slope-zero equations is computationally expensive and this check is replaced by the weaker condition that each of the five matrices $M_a = (d_{ijk}k_a^i)$ has at least one positive and one negative entry. Moreover, the same should hold for every linear combination $v^a M_a$. In practice, considering all the vectors $v^a$ with integer entries between $-2$ and $2$ provides a strong enough check. 

\item[(C4) Spectrum:] cohomology dimensions must satisfy\\[1mm]
${\bf 10}$-multiplets: $h^1(X,V)=3|\Gamma|$\\
no $\overline{{\bf 10}}$-multiplets: $h^2(X,V)=0$\\
$\overline{{\bf 5}}$-multiplets: $h^1(X,\wedge^2V)=3|\Gamma|+n_h$,  $n_h>0$\\
Higgs:  $h^2(X,\wedge^2V)=n_h$\\[1mm]
Here $|\Gamma|$ is the order of the discrete group $\Gamma$ and $n_h$ represents the number of Higgs doublet pairs. In the absence of a cohomology formula, (C4) can be replaced by the weaker constraint (C4').
\item[(C4') Chiral spectrum]
$\chi(X,V)=\chi(X,\wedge^2 V)=3|\Gamma|$
\end{description}

\begin{description}
\item[(C5) Equivariance] Require that $V$ descends to a bundle on $X/\Gamma$. 
For symmetries acting trivially on the basis $(J_1,\ldots, J_h)$ we require that the Euler characteristic of every (maximal) partial sum $\oplus_{a_i} L_{a_i}$ in~$V$ consisting of line bundles with identical first Chern classes, is divisible by $|\Gamma|$. For symmetries with a non-trivial action on the basis $(J_1,\ldots, J_h)$, $V$ must admit a partition into partial sums that are invariant under the induced action of $\Gamma$ on $(J_1,\ldots, J_h)$ and, moreover, the Euler characteristic of each partial sum must be divisible by $|\Gamma|$.
\end{description}
The GA scans discussed below have been carried out on four different Calabi-Yau threefolds realised as complete intersections in products of projective spaces. Using the standard notation for configuration matrices, with superscript indices on $X$ indicating the Hodge numbers $(h^{1,1}(X),h^{1,2}(X))$ and a subscript index indicating the position in the CICY list~\cite{Candelas:1987kf}, these four manifolds are generic members of the following deformation families:
\begin{equation}\label{eq:cicys}
\begin{aligned}
\scriptsize{X_{7862}^{(4,68)} = \cicy{\IP^1 \\ \IP^1\\ \IP^1\\ \IP^1}
{ ~2 \!\!\\[1pt]
  ~2\!\! & \\[1pt]
  ~2\!\! & \\[1pt]
  ~2\!\!}}~&,~
\scriptsize{X_{7447}^{(5,45)} = \cicy{\IP^1 \\ \IP^1\\ \IP^1\\ \IP^1\\ \IP^1}
{ ~1&1 \!\!\\[0pt]
  ~1&1\!\! & \\[0pt]
  ~1&1\!\! & \\[0pt]
   ~1&1\!\! & \\[0pt]
  ~1&1\!\!} } \\
\scriptsize{X_{5302}^{(6,30)} = \cicy{\IP^1 \\ \IP^1\\ \IP^1\\ \IP^1\\ \IP^1\\ \IP^1}
{ ~0&1&1 \!\!\\[1pt]
~0&1&1 \!\!\\[1pt]
  ~1&1&0\!\! & \\[1pt]
    ~1&1&0\!\! & \\[1pt]
  ~1&0&1\!\! & \\[1pt]
    ~1&0&1\!\!}}~&,~
\scriptsize{
X_{4071}^{(7,27)} = \cicy{\IP^1 \\ \IP^2 \\ \IP^1\\ \IP^1\\ \IP^1\\ \IP^2 \\ \IP^3}
{~1 & 1 & 0 & 0 & 0 & 0 & 0 & 0 \\[1pt]
 ~0 & 1 & 1 & 0 & 0 & 0 & 1 & 0 \\[1pt]
~ 0 & 0 & 1 & 0 & 0 & 1 & 0 & 0 \\[1pt]
~ 0 & 0 & 0 & 0 & 2 & 0 & 0 & 0 \\[1pt]
~ 0 & 0 & 0 & 1 & 1 & 0 & 0 & 0 \\[1pt]
~ 1 & 0 & 0 & 0 & 0 & 1 & 0 & 1 \\[1pt]
~ 0 & 0 & 0 & 1 & 1 & 0 & 1 & 1 } }
\end{aligned}
\end{equation}
All four embeddings are favourable, in the sense that a basis $(J_1,\ldots J_h)$ of $H^2(X,\mathbb Z)$ can be obtained by pulling back to $X$ the K\"ahler classes of the $h$ projective factors. 
Line bundle cohomology formulae on the manifolds $X_{7862}$ and $X_{7447}$, used to implement the constraints (C4) in the GA searches, are presented in Appendix~\ref{app:coh}. For the manifolds $X_{5302}$ and $X_{4071}$ cohomology formulae are not yet available and we have used the weaker spectrum constraint (C4'). The first three manifolds admit symmetries of orders $2$ and $4$ which leave the basis $(J_1,\ldots J_h)$ invariant, while $X_{4071}$, admits a free action by $\mathbb Z_2$ which maps $(J_1,J_2,J_3,J_4,J_5,J_6,J_7)\mapsto (J_1,J_6,J_3,J_4,J_5,J_2,J_7)$.


\section{The Genetic Algorithm and Quantum Annealing}
Fixing the manifold $X$, a sum of five line bundles $V$ is specified by $4h$ integers $(k_a^i)_{a=1,\ldots 4}^{i=1,\ldots ,h}$, where the condition (C1) is used to fix the fifth line bundle in terms of the first four. There are no a priori bounds on these $4h$ integers. However, our previous experience from systematic scans~\cite{Anderson:2013xka, Constantin:2018xkj} indicates that only a relatively small range is relevant, as bundles involving larger integers either violate the anomaly cancellation condition or fail to match the required Euler characteristic. We choose this range as $k_a^i\in\{-2^n+1,\ldots ,2^n\}$, so that every integer can be encoded by $n+1$ 
bits without redundancy, and a complete model is described by a bit list of length $N_{\rm bits}=4h(n+1)$. In practice, we take $n=3$ for the first three manifolds and $n=2$ for the manifold $X_{4071}$. 

The classic GA algorithm begins by forming a random population of $N_{\rm pop}$ individuals, that is by generating $N_{\rm pop}$ random binary string genotypes of length $N_{\rm bits}$. To decide how successful a particular individual is, we define a {\itshape fitness function} $f:\mathbb{F}_2^{N_{\rm bits}}\rightarrow\mathbb{R}$ on this set of binary strings, which indicates how close the corresponding bundle comes to satisfying conditions (C1)--(C5). The detailed definition of $f$ is presented in Appendix~\ref{app:fitness}. The population is then evolved via the three main evolutionary ingredients: selection, breeding and mutation. We use a selection method based on fitness-ranking, which means that  individuals are selected for breeding with a probability that increases linearly with their ranking, such that the probability for the fittest individual to be selected  is a multiple $\alpha$ of the probability for the least fit one. Typically, $\alpha$ is chosen in the range $2 \leq \alpha \leq 5$. The breeding of the $N_{\rm pop}/2$ pairs that are selected in this manner is implemented by cutting and splicing each pair at a number of matching random points. Typically (and, in particular, in this work) a single point cross-over performs well enough, in which a cut is made at a single random point and the `tails' swapped. Mutation is the final step, in which a small randomly selected fraction of bits in the newly formed generation is flipped. It is worth highlighting the crucial importance of mutation, in the absence of which the system stagnates. As an additional feature, our implementation includes {\itshape elitism}, which means that the fittest individual in every generation is copied to the next generation without modification.

The genetic quantum annealing algorithm (GQAA) described in Ref.~\cite{Abel:2022bln} makes a further step by realising the genotype of individuals in a quantum mechanical way, that is, as quantum reads on a system of spins on a quantum annealer.
This approach uses quantum annealing to enhance the GA but maintaining the same topology for the algorithm. This sidesteps the difficulty of encoding the problem directly onto the annealer (for recent discussions in the Physics context see Ref.~\cite{Abel:2022wnt} and also Ref.~\cite{Abel:toappear}). 

The manner in which such a GQAA enhances the classical GA is motivated by the way that classical GAs work. To understand this we can use the {\it schema} theorem of Holland as a rough guide (notwithstanding its still controversial status). According to the theorem, the classical GA works by propagating favourable sets of important alleles ({\it i.e.} the schema in question) throughout the population, such that the number of individuals with a good schema will grow exponentially with time. However there is clearly some redundancy in the mechanism, because the only way that the fitness gifting abilities of a particular schema can be represented is through the number of individuals in the population that carry it. The GQAA works by instead representing individuals in terms of continuous biases and couplings on a quantum annealer. These continuous allele values are called the {\it classical genotype}. In order to extract the phenotypes of all the individuals, the first step is to produce a so-called {\it quantum genotype} for them all by reading off the corresponding discrete spin values produced in a quantum anneal. The quantum genotypes that emerge from the quantum anneal are isomorphic to those in the classical GA. Thereafter the calculation of the phenotype and fitness, the selection and the breeding is all performed classically in the usual way, with the result  being used to define the next generation of biases and couplings. 

The advantage of this arrangement is that now the fitness can be represented continuously in the spin biasing of each individual. Thus, for example, the classical genotype of a very fit individual will strongly bias its preferred quantum genotype, while a weaker individual is more likely to be influenced by the stronger individuals to which it couples. In this way the represention of the fitness yielding advantage of a particular schema is enhanced beyond simply counting the number of individuals in the population that carry it. This quantum annealing step can then be thought of as a form of {\it directed mutation}, namely a mutation in which the prior fitness of the parents influences the offspring that are produced, as does the presence of much fitter individuals in the population. Indeed, it completely replaces the classical mutation step. There are several other aspects of the GQAA (especially regarding the preferred format of the couplings between individuals in the population) which are further described in  Ref.~\cite{Abel:2022bln}.

Note that in the limit in which there is no coupling between the spins on the annealer such that there are only biases, and in which the annealing is carried out perfectly adiabatically, the classical genotype determines the quantum genotype exactly, and the GQAA becomes a classical GA in this limit. This allows a direct comparison of the potential enhancement conferred by the GQAA using an otherwise identical system. 


\section{Results}
Let us begin with the classical GA. We have implemented the classic genetic algorithm and the line bundle environment (performing the binary encoding and the computation of the fitness function) in~C, and the code is available here~\cite{githubcGA,githubclb}. 
We performed $7$ different searches, as summarised in Table~\ref{table:results}. Each search was divided into a large number of genetic episodes, with every episode containing $300$ generations of $300$ individuals each. The  mutation rate was set to $0.5\%$, and the selection probability factor to $\alpha = 3$. 
\begin{center}
\begin{table}[!h]\caption{Summary of results for the $7$ GA searches. The table compares the number of models found here (GA) with numbers found in previous comprehensive searches (Scan) for manifolds with $h<7$, both as actual numbers and as percentages. For the first three manifolds these numbers refer to the models that pass a sufficient criterion for poly-stability, performed after the GA search. The last column indicates the fraction of the environment explored in the GA search.}\label{table:results}
\begin{tabular}{||c c c | c |  c  c | c  c ||}
 \hline
 Manifold & $h$& $ |\Gamma|$ & Range & GA & Scan  & Found & Explored \\ [0.5ex] 
 \hline\hline
 7862 & 4 & 2 & [-7,8] & 5  & 5  & 100\% &  $10^{-10}$\\ 
 7862 & 4 & 4 & [-7,8] & 30  & 31  & 97\% & $10^{-10}$ \\ 
 \hline
 7447 & 5 & 2  & [-7,8] & 38  & 38  & 100\% & $10^{-14}$ \\
 7447 & 5 & 4  & [-7,8] & 139  & 154  & 90\% & $10^{-14}$ \\
 \hline
 5302  &6 & 2 & [-7,8] & 403  & 442  & 93\% & $10^{-19}$  \\
 5302  &6 & 4 & [-7,8] & 722  & 897  & 80\% &  $10^{-19}$ \\
 \hline
 4071  &7 & 2 & [-3,4] & 11,937 & N/A & N/A & $10^{-14}$ \\
 \hline
\end{tabular}
\end{table}
\vspace{-12pt}
\end{center}

\subsection{The manifolds $X_{7862}$, $X_{7447}$ and $X_{5302}$}
Systematic and comprehensive scans on these manifolds have been previously carried out in Ref.~\cite{Anderson:2013xka}. On the manifold $X_{5302}$ a search using reinforcement learning was carried out in Ref.~\cite{Larfors:2020ugo}. Our purpose here is to gauge the GA performance as a heuristic method of search. The results are surprising. For the manifold $X_{7862}$ with $h^{1,1}(X)=4$, the environment contains $\sim\!\!10^{19}$ line bundle sums\footnote{The comprehensive scan of Ref.~\cite{Anderson:2013xka} on environments of this size was only possible due to the split nature of the bundle, which implied that vast regions of the solution space could be discarded by imposing constraints on individual line bundles, pairs of line bundles etc. The present GA search does not make use of such simplifications.}. All $\mathbb Z_2$-models and $97\%$ of the  $\mathbb Z_4$-models were found after visiting a fraction of $10^{-10}$ of this environment. For the manifold $X_{7447}$ with $h^{1,1}(X)=5$, the size of the environment is $\sim\!\!10^{24}$. All $\mathbb Z_2$-models and $90\%$ of the  $\mathbb Z_4$-models were found after visiting an even smaller fraction of $10^{-14}$ of the environment. Most impressively, for the manifold $X_{5302}$ with $h^{1,1}(X)=6$ the environment contains $\sim\!\!10^{29}$ bundles and after visiting only a tiny fraction of $10^{-19}$ of it,  $93\%$ of the  $\mathbb Z_2$-models and $80\%$ of the  $\mathbb Z_4$-models were found. 

\begin{figure}
	\centering
	\includegraphics[width=0.44\textwidth]{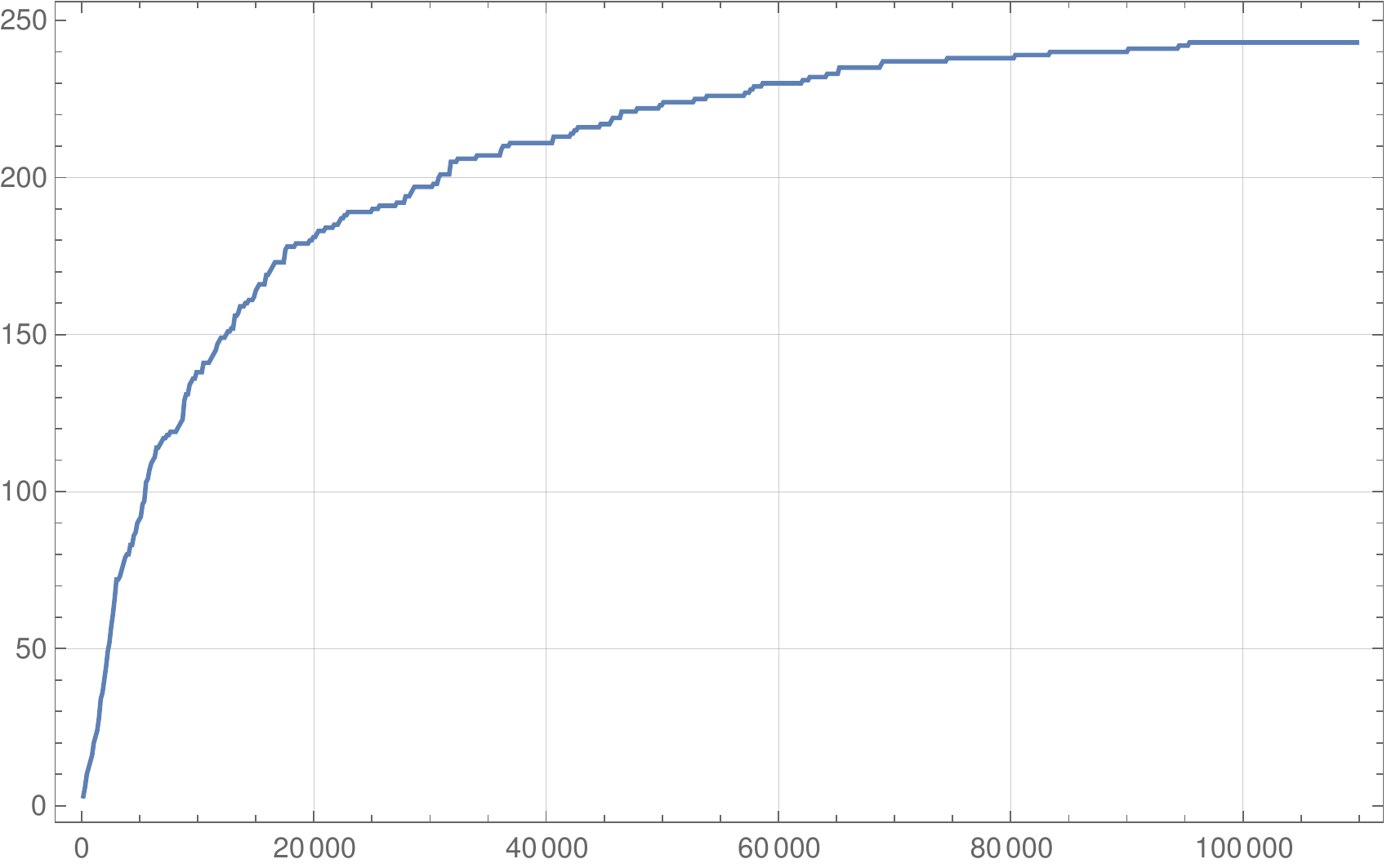}
	\caption{Saturation plot for the GA search on $X_{7447}$ with $|\Gamma|=4$ and $h^{1,1}(X_{7447})=5$. The horizontal axis represents the number of genetic episodes, in each episode a number of 90,000 states being visited. The vertical axis corresponds to the number of inequivalent models found in the search satisfying the necessary criterion (C3) for poly-stability. The computational time for a genetic episode is $O(10)$ seconds on a standard machine.}
	\label{fig:7447-4}
\end{figure}

In \fref{fig:7447-4} we present the saturation curve for the number of inequivalent $\mathbb Z_4$-models found in the GA search on $X_{7447}$ as a function of the number of states visited. Similar saturation curves were also obtained in the other cases. An important common feature of these saturation curves, relevant for evaluating the performance of the~GA, is that the initial rate of finding new viable models is of order~$1$ (inequivalent) models per 100 episodes. This implies that, although the size of the environment increases by several orders of magnitude with every additional K\"ahler parameter, while the number of viable models is expected to increase only by an order of magnitude, the initial rate at which the GA identifies these is independent of the number of K\"ahler parameters.

The computational time required for a genetic episode is $O(10)\,$s on a standard desktop and displays a linear increment with the number of K\"ahler parameters ($\sim\!\!23\,$s for $X_{5302}$, compared to $\sim\!\!17.5\,$s for $X_{7447}$ and $\sim\!\!12\,$s for $X_{7862}$). This means that each of the searches mentioned above finished within a few hours on a cluster of 100 CPUs. 
\begin{figure}
	\centering
	\includegraphics[width=0.46\textwidth]{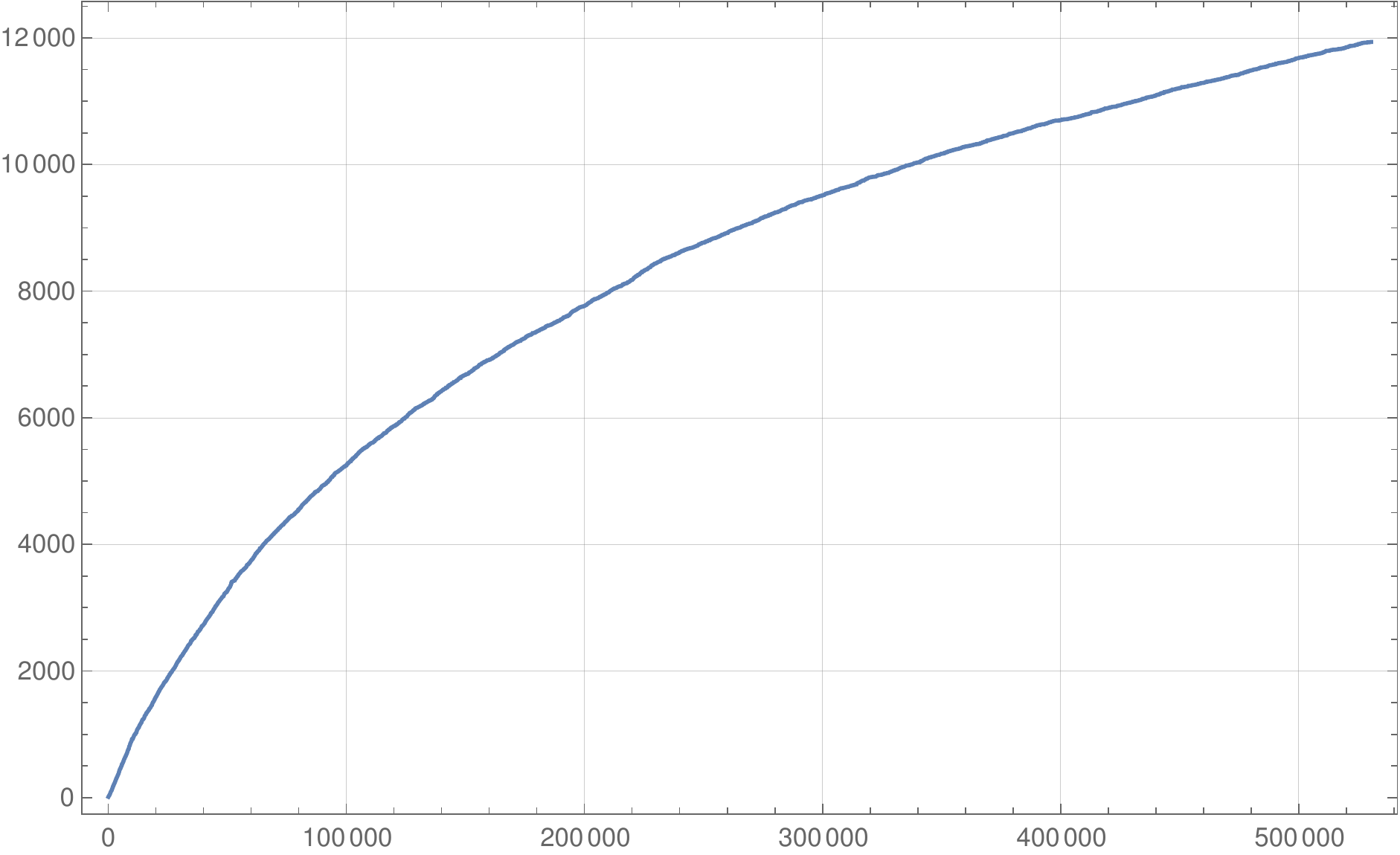}
	\caption{Saturation plot for the GA search on  $X_{4071}$ with $|\Gamma|=2$ and $h^{1,1}(X_{4071})=7$.  The horizontal axis represents the number of genetic episodes, each episode containing a number of 90,000 visited states. The vertical axis corresponds to the number of inequivalent models found in the search, satisfying the necessary criterion (C3) for poly-stability. The computational time for a genetic episode is $O(1)$ minute on a standard machine.}
	\label{fig:4071}
\end{figure}

\subsection{The manifold $X_{4071}$}
The manifold $X_{4071}$ pushes the search for realistic string models of particle physics into a new realm of larger Picard numbers. The favourable representation shown in Eq.~\eqref{eq:cicys} was taken from the maximally favourable CICY list of Ref.~\cite{Anderson:2017aux} and the $\mathbb Z_2$-symmetry from the recent classification of cyclic freely acting symmetries undertaken in Ref.~\cite{Gray:2021kax}. The group $\mathbb Z_2$ acts non-trivially on the second cohomology of the manifold, leading to a more involved bundle equivariance check as discussed in Section~\ref{sec:heterotic}. To our knowledge, this is the first instance when a symmetry of this type was considered for the purpose of a large-scale search. 

 The classification of Ref.~\cite{Gray:2021kax} gives examples of favourable CICYs with numbers of K\"ahler parameters as large as~$15$ that admit freely acting cyclic symmetries. All of these examples involve non-trivial actions on the second cohomology and the methods discussed here for the manifold $X_{4071}$ are directly applicable in these cases. Performing searches on such manifolds would very likely result in a plethora of viable string models.

The saturation curve for the number of inequivalent models found on $X_{4071}$ as a function of the number of states visited is shown in \fref{fig:4071}. The plot indicates that after 500,000 genetic episodes saturation has not been reached. However, by doubling the computational time a good degree of saturation would likely be achieved. 

Due to the more involved equivariance checks, the computational time required for a single genetic episode was slightly longer than for the previous manifolds and averaged at around 1 minute.


\subsection{Results with GQAA}
Let us now compare the potential performance of the GQAA on $X_{7447}$ and $X_{5302}$, with the results obtained using the classical GA. For quantum annealing, we used D-Wave's {\texttt Advantage\_system4.1} whose annealer contains 5627 qubits, connected in a {\itshape Pegasus} structure, with a total of $40279$ couplings between them. As such machines are still in development it is not possible at the time of writing to reproduce analogous plots to the saturation plots in Figures~\ref{fig:7447-4} and \ref{fig:4071} for GQAA. (Indeed, considering only the available space on the annealer, a GQAA reproduction of Fig.~\ref{fig:7447-4} with the same population and the same range for the integers $k_a^i$ would already require $24000$ qubits, which is far beyond the available number of qubits on the D-Wave's {\texttt Advantage\_system4.1}.) 

Given these practical constraints, comparing the GQAA with the GA then requires careful consideration. For example one might consider resorting to smaller problems, such as a saturation plot on $X_{7862}$ with $k_a^i \in [-2,1]$ using a smaller population. However such a  problem is then already somewhat trivial for both algorithms to solve since there are a high number of perfect models in the search space. In other words the classical GA  already finds a solution in every other genetic episode (by comparison with the saturation plot of the 7447 model in Fig.~\ref{fig:7447-4} where it finds a perfect model roughly once in every 100 genetic episodes), so there is little room for the GQAA to show advantage over the classical GA (although we should add that both algorithms are still orders of magnitude better than a random search).  

\begin{figure}[h!]
	\centering
	\includegraphics[width=0.4\textwidth]{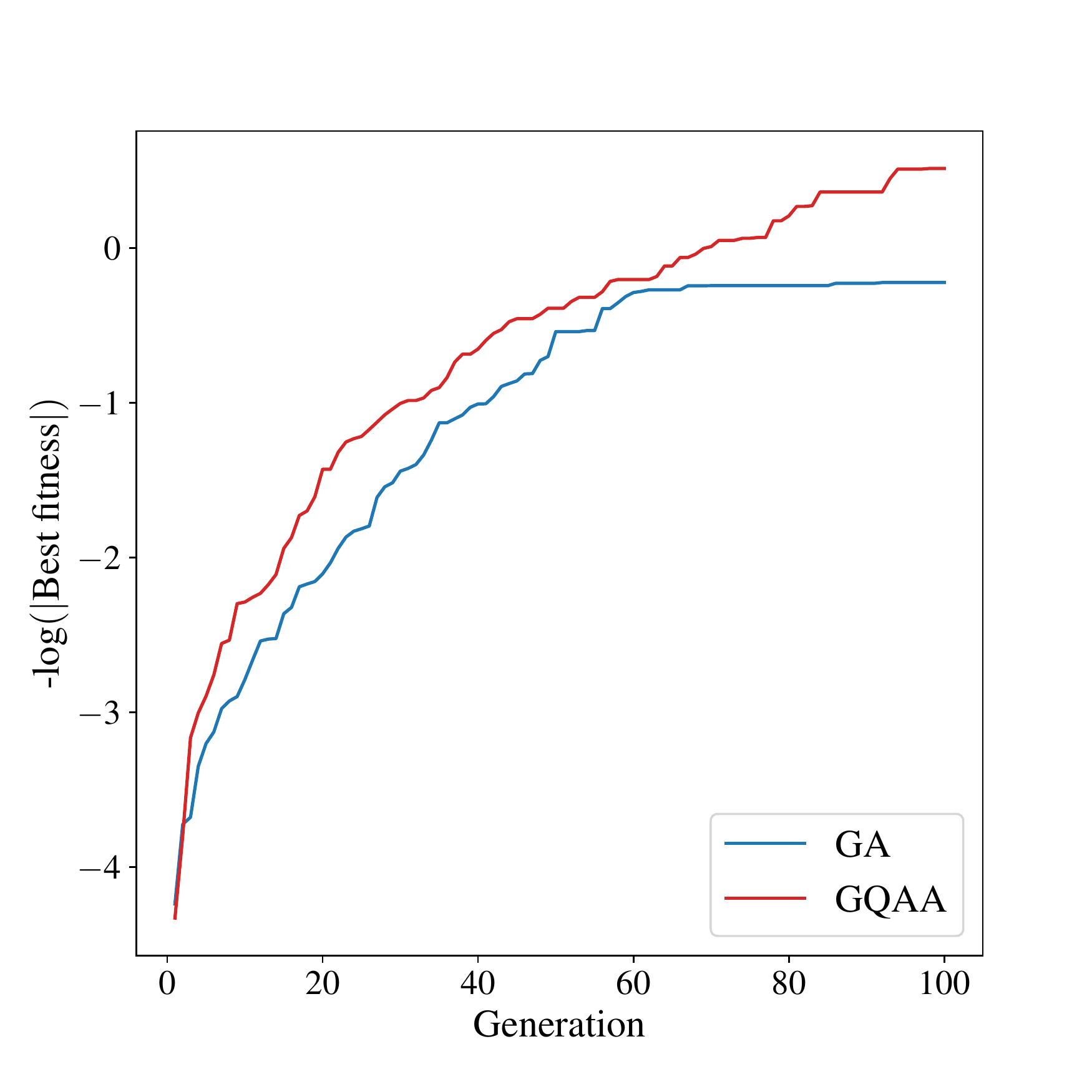}
	\caption{Fitness of the fittest individual for both GA and GQAA for the first 100 generations on $X_{7447}$ with $|\Gamma| = 4$. The optimal value of the GA mutation rate is 0.5\% and the range for the integers $k_a^i$ is chosen to be $[-4,3]$.  $N_{\rm pop}$ was set to 50 for both GA and GQAA. The fitness was averaged over 20 runs. All the other parameters related to the GQAA part are specified in Table~\ref{table:params}.}
	\label{gen_fit_7447}
\end{figure}

\begin{figure}[h!]
	\centering
	\includegraphics[width=0.4\textwidth]{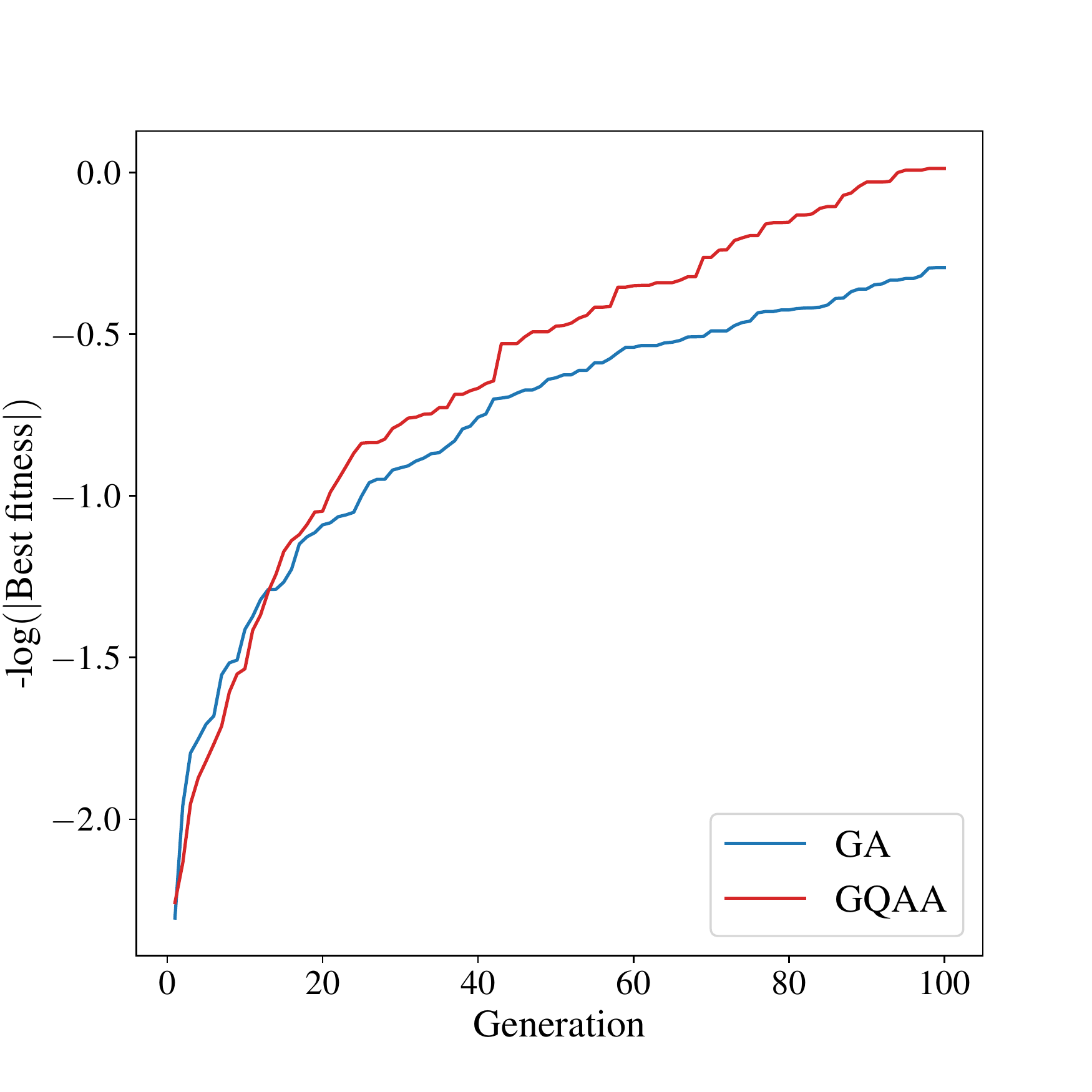}
	\caption{Fitness of the fittest individual for both GA and GQAA for the first 100 generations on $X_{5302}$ with $|\Gamma| = 2$. The optimal value of the GA mutation rate is 1\% and the line bundle integers are chosen in the range $k_a^i \in [-7,8]$.  $N_{\rm pop}$ was set to 35 for both GA and GQAA. The fitness was averaged over 20 runs. All the other parameters related to the GQAA part are specified in Table~\ref{table:params}.}
	\label{gen_fit_5302}
\end{figure}

Therefore to ensure that we are analysing a problem that is {\it hard for the traditional GA}, we can instead compare the early improvement in the best fitnesses for the much more difficult cases, and with higher $k_a^i$.  We~show this in Figures~\ref{gen_fit_7447} and \ref{gen_fit_5302}, which compare the fitness evolution for the two algorithms on $X_{7447}$ and $X_{5302}$, respectively. After optimising all the GA parameters and choosing a suitable set of GQAA parameters (which can be found in Table~\ref{table:params}), we determined the fitness of the fittest individual for the first $100$ generations for both GA and GQAA, averaged over 20 runs. 

By this measure we can indeed see evidence that the GQAA has advantage over the classical GA. We note that the GQAA best fitness grows faster throughout the generations than that of the classical GA. Indeed, after $100$ generations, the GQAA best fitness is, on average, $\sim52 \%$ better than GA in the first case and $\sim26 \% $ in the second case, respectively. We also note a much smoother behaviour of the fitness improvement in the GQAA. For example, the stall in fitness improvement for the classical GA on the $X_{7447}$ manifold is reproducible and remains after many more runs have been performed. Thus, it appears that, depending on the manifold in question, the GA can encounter blocks in the fitness improvement that the QGAA is able to circumvent. 
\begin{table}[h!]
\begin{center}
\resizebox{8.7cm}{!}
{
\begin{tabular}{| c | c | c |  }
 \hline
  \textbf{Parameter} & \textbf{Description} & \textbf{Value} \\ \hline \hline 
  {Topology} & Polyandric {$J_{\ell m}$ couplings} &  `Islands' \\
 \hline
 $\alpha $ & Selection probability factor & 4.0 | 1.16 \\
 \hline
 $\alpha_p $ & Nepotism & 0.05 | 0.6 \\
  \hline
 $\rho $ & Proportion of antiferromagnetic&  0.5 \\
  \hline
 $\rho' $ & Proportion of enhanced couplings &  6.4\% \\
  \hline
 $\kappa $ & Strength of enhanced couplings & $-~\alpha\, \times \alpha_p$ \\
  \hline
 $s_q $ & Minimum anneal parameter & 0.75 | 0.2  \\
  \hline
 $J_{ij} $ & Coupling strength & $\pm 0.08   | \pm 0.15$ \\
  \hline
\end{tabular}
}
\end{center}
\caption{GQAA parameters and related values. For definitions, see Ref.~\cite{Abel:2022bln}. Two entries in the ``Value'' column refer to $X_{7447}$ and $X_{5302}$, respectively. A single entry refers to both manifolds.}
\label{table:params}
\end{table}

This partly explains why the improvement on the classical GA in the first case (Fig.~\ref{gen_fit_7447}) is twice that in the second case (Fig.~\ref{gen_fit_5302}): 52\% and 26\%, respectively. However, we should also note in this respect that besides being intrinsically dependent on the characteristics of the problem (manifold structure, range of the variables, {\it etc}.), the efficacy of the GQAA depends strongly on the choice of GQAA parameters in Table~\ref{table:params}. Thus while for the GA it is possible to optimise meta-parameters such as mutation rate, this becomes too time-consuming an operation for the GQAA due among other things to the much larger number of meta-parameters (the values of all the couplings for example). Thus we consider Figures~\ref{gen_fit_7447} and \ref{gen_fit_5302} to be evidence of advantage even before a full optimisation has been performed. 
From these results, it seems reasonable to believe that once the technological limitations have been  overcome, a GQAA saturation plot on the manifold $X_{7447}$ would require roughly half of the genetic episodes required by the classical GA (Fig.~\ref{fig:7447-4}) to reach saturation or possibly even less once a full optimisation becomes possible. 

Finally it is worth mentioning a salient fact which is that one might suppose that the quantum annealing step  could be replaced with simulated thermal annealing. However in contrast to quantum annealing, thermal annealing appears to behave differently such that we were not able to find any choice of meta-parameters for which it could offer any benefit.

\section{Conclusion}\label{sec:Conclusion}

The main lesson emerging from the present work is that the size of the string landscape is no longer a major impediment in the way of constructing realistic string models of Particle Physics. Using genetic algorithms we were able to scrutinise spaces of string compactifications of sizes as large as $\sim\!\!10^{30}$ and identify the vast majority of good solutions residing in these spaces after visiting only a tiny fraction of the total number of states. This has been carried out for heterotic line bundle models on Calabi-Yau threefolds with $4, 5, 6$ and $7$ K\"ahler parameters but the basic methodology applies to other string constructions as well. We also presented evidence that the method can be enhanced with quantum annealing,

Extending the methods to manifolds with larger numbers of K\"ahler parameters (for example the favourable CICYs of Ref.~\cite{Gray:2021kax} which include examples with up to $15$ K\"ahler parameters) is perfectly achievable. While finding all the models that satisfy the conditions (C1)--(C5) of Section~\ref{sec:heterotic} may be impossible on such manifolds (the expected numbers of viable models being too large even to store), our results suggest that GAs provide the means to produce string models on demand. The method can also be extended to manifolds from the Kreuzer-Skarke list~\cite{Kreuzer:2000xy}, provided that more examples with a non-trivial fundamental group are found~\cite{Braun:2017juz}.

In order to impose the full spectrum constraints (as we did here for the manifolds $X_{7862}$ and $X_{7447}$), explicit line bundle cohomology formulae need to be obtained, using a combination of algebro-geometric methods~\cite{Brodie:2021toe, Brodie:2021nit, Brodie:2021zqq} and machine learning techniques~\cite{Klaewer:2018sfl, Brodie:2019dfx, Bies:2020gvf}. 

Finally, it is important to stress that the constraints (C1)--(C5) of Section~\ref{sec:heterotic} do not represent a complete list of requirements, but only lead to a broad brush version of the Standard Model. The formidable success of GAs and the immediate access to cohomology data provided by the line bundle cohomology formulae offer the possibility of substantially refining the requirements. One of the many ways in which this can be accomplished would proceed by first identifying a number of Froggatt-Nielsen models with four flavour $U(1)$ symmetries that can explain the observed hierarchies of fermion masses and mixings. The successful Froggatt-Nielsen models would then correspond to specific $U(1)$-charge assignments for the bundle moduli fields, measured by $H^2(X,V\otimes V^\ast)$, which would be searched for along with imposing the other constraints. Other requirements that can be included are the absence of fast proton decay operators, as well as various model-dependent constraints for neutrino physics. Such additional requirements will lead to a large reduction in the number of viable models, making complete (and more targeted) GA searches possible even for models with a larger number of K\"ahler parameters.


\section*{Acknowledgements}
The research activities of SAA were supported by the STFC grant ST/P001246/1. AC's research was supported by a Stephen Hawking Fellowship, EPSRC grant $\text{EP/T016280/1}$. TRH was supported by an STFC studentship. 

\vspace{30pt}


\bibliography{bibliography2}
\bibliographystyle{inspire}


\appendix

\section{The fitness function}\label{app:fitness}
The fitness function $f(X,V)$ is the measure of how close a line bundle sum $V$ over a smooth Calabi-Yau threefold $X$ comes to satisfy the constraints (C1)--(C5) of Sec.~\ref{sec:heterotic}. It receives several contributions, 
\begin{equation}
\begin{aligned}
f &= f_{\rm anom} + f_{\rm ind} + f_{\rm slope}+ f_{\rm equiv}+ f_{\rm str.\,gp}+f_{\rm spec}\; ,
\end{aligned}
\end{equation}
which we now discuss in turn. The contribution associated with the cancellation of anomalies is
\begin{equation}
f_{\rm anom} = 10 \sum_{i=1}^{h} \frac{\min(c_{2,i}(V)-c_{2,i}(TX),0)}{h k_{\rm max}^2\rk(V)}~,
\end{equation}
where $k_{\rm max}=2^{n}$ is the (absolute) maximal line bundle integer allowed in the search. The sum contains $h$ terms that are quadratic in the line bundle integers, hence the pre-factor $(hk_{\rm max}^2)^{-1}$. The factor of $10$ and the numerical factors appearing below in the expressions for the other contributions to the fitness function are arranged such that  all contributions are of roughly the same order of magnitude for a typical bundle $V$.

 The contribution from the Euler characteristic of $V$ is 
\begin{equation}\label{find}
f_{\rm ind} = -100\frac{|\ind(V) + 3|\Gamma| |}{h k_{\rm max}^3 \rk(V)}~.
\end{equation}
The necessary slope-0 checks discussed under (C3) involve a number of matrices which are required to have both positive and negative entries. If a number $n_{\rm pos}$ of these matrices are found to have non-negative entries only and  a number $n_{\rm neg}$ are found to have non-positive entries only, there is a (negative) contribution to the fitness function equal to
\begin{equation}
f_{\rm slope} = - \frac{n_{\rm pos}+n_{\rm neg}}{10}~.
\end{equation}
For equivariance, in the case of symmetries acting trivially on the second cohomology of $X$ we have a contribution 
\begin{equation}\label{equiv1}
f_{\rm equiv}=-\!\!\!\!\!\!\!\!\! \sum_{\text{distinct } L\subset V} m(L) \chi(X,L) \text{ mod } |\Gamma| ~,
\end{equation}
where $m(L)$ is the number of times $L$ appears in the line bundle sum $V$.

Symmetries with a non-trivial action $\Gamma$ on the second cohomology permute non-trivially the projective space factors in the embedding of $X$, which amounts to a permutation of line bundle integers in $V$. The fitness contribution from equivariance then has to take into account two aspects. On the one hand, in the ideal case the permutation induced by $\Gamma$ on $V$ will produce a bundle that can be identified with $V$ up to re-orderings of the line bundles. The failure to achieve this is measured by summing over the absolute values of the differences between the line bundle integers in $V$ and the integers obtained after applying the $\Gamma$-permutation and a line bundle re-ordering, and minimising over all possible re-orderings. We call this minimal sum $M$, which in the ideal case vanishes. Furthermore, provided that $M=0$, we compute an equivariance contribution analogous in spirit to \eqref{equiv1} above, with the difference that we first partition the line bundles in $V$ into parts that are formed from the cycles of the $\Gamma$-permutation, compute their Euler characteristics, mod out by $|\Gamma|$ and sum over all the parts. We denote this value as $N$.  With these considerations, the total fitness contribution from equivariance is taken to be 
\begin{equation}\label{eqn:equiv}
    f_{\rm equiv} = \begin{cases}
        - |\Gamma|-M &  M\neq 0 \\
         -N & M=0
    \end{cases},
\end{equation}
where in the case $M\neq 0$ we have added a default penalty of $-|\Gamma|$, corresponding to the maximal penalty that can be accrued from $N$ when $M=0$. This default penalty provides an incentive to evolve towards achieving $M=0$.

The contribution from the constraint on the structure group is given by
\begin{equation}
f_{\rm equiv}=-\frac{|S|}{10}~,
\end{equation}
where $S$ is the collection of subsets of line bundles in $V$ whose sum of first Chern class vanishes. 

The contribution corresponding to the spectrum, on manifolds where a cohomology formula is available, is computed as
\begin{equation}
\begin{aligned}
f_{\rm spec}&{=}-1000 \frac{h^0(X, V) {+} h^3(X, V) {+} h^0(X, \wedge^2 V) {+} h^3(X, \wedge^2 V)}{hk_{max}^3 \rk(V)^2}\\
& -100 \frac{h^2(X,V)}{h k_{max}^3}+\frac{\theta(h^2(X, \wedge^2 V)-1/2)-1}{10}\\
&-5\frac{\max(h^2(X,\wedge^2 V)/|\Gamma| -2,0)}{h k_{max}^3 \rk(V)}~,
\end{aligned}
\end{equation}
where the first term corresponds to the requirement that the zeroth and the top cohomologies of $V$ have to vanish in view of slope-stability, while the other terms correspond to the exact spectrum constraints in (C4). Concretely, the second term corresponds to the absence of $\overline{\mathbf{10}}$-multiplets, the third term penalises the lack of Higgs multiplets, while the fourth term penalises the presence of more than two pairs of Higgs doublets. There is no further contribution from the $\overline{\mathbf{10}}$-multiplets and the $\overline{\mathbf{5}}$-multiplets, given the $f_{\rm ind}$ contribution in \eqref{find} above. In the absence of a cohomology formula we set $f_{\rm spec}=0$, since the index constraint (C4') has already been  taken care of in \eqref{find}.

\section{Line Bundle Cohomology Formulae}
\label{app:coh}
\subsection{The manifold $X_{7862}$}
Cohomology formulae for the tetra-quadric manifold $X_{7862}$, which corresponds to a generic hypersurface of multi-degree $(2,2,2,2)$ in $(\mathbb P^1)^{\times 4}$, have previously been given in Refs.~\cite{Constantin:2018otr,Buchbinder:2013dna, Constantin:2018hvl}. However, these earlier formulae were only correct in a finite range of line bundle integers. A~complete formula has appeared in Ref.~\cite{Constantin:2022jyd}, and here we follow the arguments of this paper. For simplicity, in this section we write $X$ instead of $X_{7862}$. 

The embedding is favourable and also K\"ahler favourable. We denote by $(J_1,\ldots, J_4)$ the generators of the K\"ahler cone $\mathcal K(X)$ inherited from the ambient space. A line bundle $L$ over $X$ with first Chern class $c_1(L) = \sum_{i=1}^4k_iJ_i$ has Euler characteristic 
\begin{equation}\label{eq:7862_index}
\chi(X,L) = \int_X {\rm ch}(L)\cdot {\rm td}(X) =2\sum_{i=1}^4\left(k_i+\prod_{j\neq i}k_j\right)\; .
\end{equation}
Apart from $\mathcal K(X)$, the effective cone of $X$ includes an infinite number of simplicial cones. They correspond to the K\"ahler cones of isomorphic Calabi-Yau threefolds which can be reached from $X$ by a sequence of flops (see Refs.~\cite{Brodie:2021toe}). These additional cones are obtained from the K\"ahler cone by the action of an infinite group generated by the matrices
\begin{equation}
\begin{aligned}
&{\scriptsize M_1 {=} \left(\begin{array}{rrrr}{\!\!\!\!-1}&{0}&{0}&{0}\!\!\\{2}&{1}&{0}&{0}\!\!\\{2}&{0}&{1}&{0}\!\!\\{2}&{0}&{0}&{1}\!\!\end{array}\right)~,~
M_2 {=} \left(\begin{array}{rrrr}{\!\!1}&{2}&{0}&{0}\!\!\\{\!\!0}&{\!\!\!\!{-}1}&{0}&{0}\!\!\\{\!\!0}&{2}&{1}&{0}\!\!\\{\!\!0}&{2}&{0}&{1}\!\!\end{array}\right)~,~}\\[4pt]
&{\scriptsize M_3 {=} \left(\begin{array}{rrrr}{\!\!1}&{0}&{2}&{0}\!\!\\{\!\!0}&{1}&{2}&{0}\!\!\\{\!\!0}&{0}&{\!\!\!\!{-}1}&{0}\!\!\\{\!\!0}&{0}&{2}&{1}\!\!\end{array}\right)~,~~
M_4 {=} \left(\begin{array}{rrrr}{\!\!1}&{0}&{0}&{2}\!\!\\{\!\!0}&{1}&{0}&{2}\!\!\\{\!\!0}&{0}&{1}&{2}\!\!\\{\!\!0}&{0}&{0}&{\!\!\!\!{-}1}\!\!\end{array}\right)~.}
\end{aligned}
\end{equation}
Consequently, any effective line bundle $L$ is related to a line bundle $L'$ contained in the closure of the K\"ahler cone by a finite number of transformations
\begin{equation}\label{7862coh}
c_1({L'})=M_{i_1}M_{i_2}\ldots M_{i_k}c_1(L) \in \overline{\cK(X)}~.
\end{equation}
However, $h^0(X,L)=h^0(X,L')= \chi(X,L')$, since the number of global sections of a line bundle is invariant under flops and the second equality holds by Kodaira's vanishing theorem and the Kawamata-Viehweg vanishing theorem (the latter required on the walls separating the K\"ahler cone of $X$). In fact, there are a number of two-faces of $\overline{\cK(X)}$ which are not covered by the Kawamata-Viehweg vanishing theorem. These correspond to line bundles for which at least two of the integers $k_i$ vanish and the remaining integers are non-negative. In these cases, the zeroth cohomology function is simply $\prod_{i=1}^4(1+k_i)$, which can be easily traced back to the zeroth cohomology of two line bundles on $\mathbb P^1\times \mathbb P^1$. 

This procedure gives an extremely efficient method for computing the zeroth cohomology of line bundles on the tetra-quadric threefold. In practice only a small number of transformations arise in Eq.~\eqref{7862coh}, since the cones are increasingly thin as one moves away from $\mathcal K(X)$ and contain line bundles where at least one of the integers is very large.

Once the zeroth cohomology is known, the third cohomology follows by Serre duality, 
\begin{equation}
h^3(X,L) = h^0(X,L^*)~.
\end{equation}
Note that since the effective cone is convex there are no line bundles, except for the trivial line bundle, that have both $h^0(X,L)$ and $h^3(X,L)$ non-vanishing.

The middle cohomologies are related to the zeroth and the third cohomologies by the formula
\begin{equation}\label{eq:EulerCh}
h^1(X,L)-h^2(X,L) = h^0(X,L)-h^3(X,L)-\chi(X,L)~.
\end{equation}
On the tetra-quadric manifold it turns out that almost all line bundles either have $h^1(X,L)=0$ or $h^2(X,L)=0$. In all these cases Eq.~\eqref{eq:EulerCh} provides a formula for the middle cohomologies. The exceptions correspond to line bundles for which two of the line bundle integers are zero and the other two have opposite sign and are greater than $1$ in modulus. If $k_A$ and $k_B$ denote these non-zero integers, then the relation 
\begin{equation}
h^1(X,L)+h^2(X,L) = -2(1+k_A k_B)~,
\end{equation}
holds in all of the exceptional cases. Together with Eq.~\eqref{eq:EulerCh}, this fixes the middle cohomologies. 

\subsection{The manifold $X_{7447}$}
This manifold corresponds to the intersection of two generic hypersurfaces of degree $(1,1,1,1,1)$ in $(\mathbb P^1)^{\times 5}$. The line bundle cohomology structure is very similar to that of the manifold $X_{7862}$. The K\"ahler cone is five dimensional and is inherited from the embedding space. Additionally, the effective cone contains infinitely many cones obtained from the K\"ahler cone by the action of an infinite group generated by the matrices
\begin{equation}
\begin{aligned}
&{\scriptsize M_1 {=} \left(\begin{array}{rrrrr}{\!\!\!\!-1}&{0}&{0}&{0}&{0}\!\!\\{1}&{1}&{0}&{0}&{0}\!\!\\{1}&{0}&{1}&{0}&{0}\!\!\\{1}&{0}&{0}&{1}&{0}\!\!\\{1}&{0}&{0}&{0}&{1}\!\!\end{array}\right)~,~
M_2 {=} \left(\begin{array}{rrrrr}{\!\!1}&{1}&{0}&{0}&{0}\!\!\\{\!\!0}&{\!\!\!\!{-}1}&{0}&{0}&{0}\!\!\\{\!\!0}&{1}&{1}&{0}&{0}\!\!\\{\!\!0}&{1}&{0}&{1}&{0}\!\!\\{0}&{1}&{0}&{0}&{1}\!\!\end{array}\right)~,~}\\[4pt]
&{\scriptsize M_3 {=} \left(\begin{array}{rrrrr}{\!\!1}&{0}&{1}&{0}&{0}\!\!\\{\!\!0}&{1}&{1}&{0}&{0}\!\!\\{\!\!0}&{0}&{\!\!\!\!{-}1}&{0}&{0}\!\!\\{\!\!0}&{0}&{1}&{1}&{0}\!\!\\{0}&{0}&{1}&{0}&{1}\!\!\end{array}\right)~,~~
M_4 {=} \left(\begin{array}{rrrrr}{\!\!1}&{0}&{0}&{1}&{0}\!\!\\{\!\!0}&{1}&{0}&{1}&{0}\!\!\\{\!\!0}&{0}&{1}&{1}&{0}\!\!\\{\!\!0}&{0}&{0}&{\!\!\!\!{-}1}&{0}\!\!\\{0}&{0}&{0}&{1}&{1}\!\!\end{array}\right)~,~}\\[4pt]
&{\scriptsize ~~~~~~~~~~~~~~~~M_5 {=} \left(\begin{array}{rrrrr}{\!\!1}&{0}&{0}&{0}&{1}\!\!\\{\!\!0}&{1}&{0}&{0}&{1}\!\!\\{\!\!0}&{0}&{1}&{0}&{1}\!\!\\{\!\!0}&{0}&{0}&{1}&{1}\!\!\\{0}&{0}&{0}&{1}&{\!\!\!\!{-}1}\!\!\end{array}\right)~.
}
\end{aligned}
\end{equation}
As before, the zeroth cohomology dimensions of most line bundles can be obtained using the invariance under this group action and the Kawamata-Viehweg vanishing theorem. The line bundles on the boundary of $\overline{\cK(X)}$ that are not covered by the Kawamata-Viehweg theorem have at least two of the integers $k_i$ vanishing, in which case the zeroth cohomology function is simply $(1+k_A)(1+k_B)(1+k_C)$  and the other two integers, denoted by $k_A$ and $k_B$, are non-negative, which we denote by $k_A$ and $k_B$. In these cases, the zeroth cohomology function is given by $\prod_{i=1}^5(1+k_i)$. 

For the computation of higher cohomologies we use Serre duality $h^3(X,L) = h^0(X,L^*)$ and the observation that the only line bundles which have both a non-vanishing first and second cohomology  have three vanishing integers $k_i$ and the other two have opposite sign and are greater than $1$ in modulus. As before, denoting by $k_A$ and $k_B$ the non-zero integers, the following simple relation holds
\begin{equation}
h^1(X,L)+h^2(X,L) = -2(1+k_A k_B)
\end{equation}
in all of these exceptional cases. Together with Eq.~\eqref{eq:EulerCh}, this fixes the middle cohomologies. 

\end{document}